\documentclass[aps,superscriptaddress,amsmath,showpacs,twocolumn,prl]{revtex4}

\usepackage{graphicx}
\usepackage{amsfonts}
\usepackage{amssymb}
\usepackage{amsbsy}
\usepackage{amsmath}
\usepackage{latexsym}
\usepackage{natbib}
\usepackage{bm}
\usepackage{subfigure}
\usepackage{color}
\usepackage{hyperref}
\usepackage{lscape}
\def\be {\begin{equation}}
\def\ee  {\end{equation}}
\def\bea {\begin{eqnarray}}
\def\eea {\end{eqnarray}}

\newcommand*{\M}{M_{\star}}

\DeclareMathOperator{\Realpart}{Re}

\renewcommand*{\k}{\mathbf{k}}
\newcommand*{\x}{\mathbf{x}}

\DeclareMathOperator{\artanh}{artanh}
\DeclareMathOperator{\arctanh}{artanh}

\begin{document}

\title{Low energy Lorentz violation from modified dispersion at high energies}

\author{Viqar Husain}
\affiliation{Department of Mathematics and Statistics, University
of New Brunswick, Fredericton, NB, Canada E3B 5A3} 

\author{Jorma Louko}
\affiliation{School of Mathematical Sciences, University of Nottingham, 
Nottingham NG7 2RD, United Kingdom}


\date{December 2015}

\begin{abstract}

Many quantum theories of gravity propose Lorentz violating dispersion relations of the form 
$\omega = |\k|\, f\bigl(|\k|/\M\bigr)$, with recovery of approximate
Lorentz invariance at energy scales much below~$\M$.  
We show that a quantum field with this dispersion 
predicts drastic low energy Lorentz violation in atoms modelled as Unruh-DeWitt detectors, 
for any $f$ that dips below unity somewhere.  
As an example, we show that polymer quantization motivated by loop 
quantum gravity predicts such Lorentz violation below current ion collider rapidities.

\end{abstract}

\maketitle

\textit{\bfseries{Introduction.\thinspace---}}
Although local Lorentz invariance is a pillar of modern physics, 
there exist serious proposals emerging from quantum theories of spacetime
that Lorentz invariance may be 
broken at sufficiently 
high energies~\cite{AmelinoCamelia:2008qg}. 
The challenge for such proposals is twofold:  to recover approximate Lorentz invariance 
in the well tested low energy regime, and to make predictions through which 
the violation could conceivably be observed. 

If a Lorentz violating theory contains an explicit energy scale 
$\M$ that characterizes the violation, such as the Planck energy, 
it might seem straightforward to secure approximate 
Lorentz invariance at energies much below~$\M$. 
However, in quantum field theory (QFT), the effective 
low energy theory can reveal unexpected imprints 
of the theory's high energy structure~\cite{Polchinski:2011za}. 
For example, a Lorentz violation that is superficially associated 
with a large energy scale can become observable at arbitrarily 
low energies via the renormalization of a logarithmically divergent 
loop diagram~\cite{Collins:2004bp}.  
There is a similar result for the Casimir effect~\cite{bachmann-kempf}. 

The purpose of this paper is to show that such an unexpected 
low energy imprint occurs in the response of a simplified atom, 
modeled as an Unruh-DeWitt (UDW) detector~\cite{PhysRevD.14.870,DeWitt-detector}, 
coupled to a class of Lorentz violating quantum field theories in 
four-dimensional Minkowski spacetime. 

The theories we consider have a dispersion relation which 
in the preferred frame reads 
\begin{align}
\omega_{|\k|} = |\k| \,  f\bigl(|\k|/\M\bigr),  
\label{eq:disprel-gen}
\end{align}
where $\k$ is the spatial momentum, 
$\omega_{|\k|}$ is the corresponding energy, the positive constant 
$\M$ is the energy scale of Lorentz violation, 
and $f$ is a smooth positive function on the positive real line, 
with the property that $f(x)\to1$ as $x\to0_+$. 
(We set $c = \hbar = 1$.) 
This dispersion relation is approximately Lorentz invariant for 
$|\k|/\M \ll 1$. 

Despite this low energy recovery of Lorentz invariance, 
we find that, for any $f$ that dips below unity somewhere, 
transitions in an inertial UDW detector are strongly Lorentz violating  
at arbitrarily low transition energies. If
$f_c := \inf f$ satisfies $0<f_c<1$, 
the signature of the violation is a sudden enhancement in the
rate of spontaneous de-excitations and a sudden emergence of
spontaneous excitations when the detector's rapidity exceeds 
the critical value $\beta_c := \artanh(f_c)$.
Crucially, $\beta_c$ is independent of~$\M$.

As an application, we show that this class of Lorentz violating theories includes a scalar 
field quantized in the polymer quantization framework 
\cite{Ashtekar:2002sn,Ashtekar:2002vh,Husain:2010gb,Hossain:2010eb,Hossain:2009ru,Seahra:2012un}, 
a method motivated by the loop quantum gravity (LQG) program~\cite{Rovelli-book,Thiemann-book}.  
For the polymer quantized scalar field, we find that large Lorentz 
violation occurs above the critical rapidity $\beta_c \approx 1.3675$, 
a value well below $\beta \approx 3$ attained by ions at the 
Relativistic Heavy Ion Collider (RHIC)~\cite{RHIC-ref}. 
Given the close correspondence between the Unruh-DeWitt detector and 
an atom interacting with the electromagnetic field~\cite{PhysRevD.87.064038,PhysRevA.89.033835}, 
the absence of observations of Lorentz violation at the RHIC 
therefore provides a strong constraint on the possible families of modified dispersion relations.

\textit{\bfseries{Field theory with modified dispersion relation.\thinspace---}}
We consider a real scalar field $\phi$ that admits a decomposition 
into spatial Fourier modes such that 
a mode with spatial momentum $\k \ne \mathbf{0}$ is a harmonic 
oscillator with the angular frequency $\omega_{|\k|}$~\eqref{eq:disprel-gen}, where
$f$ is a smooth positive function on the positive real line. 
To allow sufficient generality, we will for the moment 
leave the small argument behaviour of $f$ unspecified, and 
we include in the Fourier 
decomposition the density-of-states weight factor
\begin{align}
\rho_{|\k|} = 
d\bigl(|\k|/\M\bigr) {\bigl[(2\pi)^3 |\k|\bigr]}^{-1/2}
 ,  
\label{eq:density-gen}
\end{align}
where $d$ is a smooth complex-valued 
function on the positive real line. 
If $f(x)=1$ and $d(x) = 2^{-1/2}$, the 
field is the usual massless scalar field. 

The central object that will be needed below is the Wightman function 
in the Fock vacuum. It is given by 
\begin{align}
G(t, \x; t', \x') 
= \int d^3\k \, | \rho_{|\k|}|^2 \, 
{e^{i {\k}\cdot(\x-\x') 
-i \omega_{|\k|} (t-t'- i \epsilon)}} , 
\label{eq:gen-propagator}
\end{align}
where the distributional character is encoded in the limit $\epsilon\to0_+$. 

\textit{\bfseries{Model atom: UDW detector.\thinspace---}}
We probe the field with a linearly-coupled 
two-level UDW detector~\cite{PhysRevD.14.870,DeWitt-detector}. This detector
model captures the essential features of an atom interacting with the 
electromagnetic field~\cite{PhysRevD.87.064038,PhysRevA.89.033835}, 
and it has been widely used to analyze motion effects in quantum field theory 
(for recent reviews see  
\cite{RevModPhys.80.787,hu-lin-louko,martinmartinez-menicucci-review}). 

The detector is pointlike and moves on the worldline ${\sf{x}}(\tau)$
where $\tau$ is the proper time. The coupling strength is
proportional to the switching function~$\chi(\tau)$, 
which is non-negative and smooth, and falls off
sufficiently rapidly at $\tau \to \pm\infty$. In first-order
perturbation theory, the probability of the detector to make a
transition from the state with energy $0$ to the state with energy
$\Omega$ (which may be positive or negative) 
is then proportional to the response function, 
$\mathcal{F}(\Omega) = \int d\tau \, d\tau' \, 
\chi(\tau) \chi(\tau') \, 
e^{-i\Omega(\tau-\tau')} 
\mathcal{W}(\tau,\tau')$, 
where $\mathcal{W}$ is the pullback of the scalar field's Wightman
function to the detector's worldline. 

When both the trajectory and the quantum state of the field are
stationary, 
$\mathcal{W}(\tau,\tau')$ depends on its arguments
only through the difference $\tau-\tau'$, and 
we may convert $\mathcal{F}$ into  
the transition rate per unit time by passing 
to the limit of adiabatic switching and 
factoring out the effective total
duration of the detection. While this procecure 
is subtle~\cite{hu-lin-louko,satz-transrate,louko-satz-curved}, for
the present purposes 
we may consider the specific switching function family 
$\chi(\tau) = \pi^{-1/4} \sigma^{-1/2} \exp\bigl[- \tau^2/(2\sigma^2)\bigr]$, 
where the positive constant $\sigma$ is the effective duration of the
interaction, and the 
normalization factor
$\sigma^{-1/2}$ provides
the conversion from transition probability to transition rate. 
For finite $\sigma$ the transition rate is given by 
$\mathcal{F}(\Omega) = \int_{-\infty}^{\infty} ds 
\, e^{-s^2/(4 \sigma^2)} \, 
e^{-i\Omega s} \, 
\mathcal{W}(s,0)$. Passing to the limit in which $\sigma$ is large compared with $1/|\Omega|$
and with the timescales over which $\mathcal{W}$ varies, we obtain 
\begin{align}
\mathcal{F}(\Omega) = \int_{-\infty}^{\infty} ds 
\, 
e^{-i\Omega s} \, 
\mathcal{W}(s,0) . 
\label{eq:transrate-nosigma}
\end{align}
We shall use equation \eqref{eq:transrate-nosigma} in the analysis that follows.

\textit{\bfseries{Inertial detector.\thinspace---}} 
We consider a detector on the inertial worldline
$\bigl(t(\tau), \x(\tau) \bigr) 
= \bigl(\tau \cosh\beta, 0, 0, \tau\sinh\beta \bigr)
$, 
where $\beta$ is the rapidity with respect to the distinguished 
inertial frame. 
For presentational simplicity we proceed assuming $\beta>0$, 
but it can be verified by a separate analysis 
that the $\beta=0$ transition rate is equal to the 
$\beta\to0$ limit of the results given below. 

From \eqref{eq:gen-propagator}
and \eqref{eq:transrate-nosigma} we
obtain 
\begin{align}
\mathcal{F}(\Omega) &= 
\int_{-\infty}^{\infty} ds 
\int d^3\k \, 
|\rho_{|\k|}|^2 \, 
e^{-i(\Omega + \omega_{|\k|} \cosh\beta- k_3 \sinh\beta)s}
\notag
\\
&= \frac{4\pi}{\sinh\beta} 
\int_0^\infty dK \, K \, 
|\rho_K|^2 
\notag
\\
& \hspace{3ex}
\times 
\int_{-\infty}^{\infty} ds \, \frac{\sin(Ks \sinh\beta)}{s}
\, e^{-i(\Omega + \omega_{K} \cosh\beta)s}
\notag
\\
&= \frac{4 \pi^2}{\sinh\beta} 
\int_0^\infty dK \, K \, 
|\rho_K|^2 
\notag
\\
& \hspace{3ex}
\times 
\Theta\bigl(K\sinh\beta - |\Omega + \omega_K \cosh\beta | \bigr) 
\notag
\end{align}
\begin{align}
&= 
\frac{\M}{2\pi \sinh\beta} 
\int_0^\infty dg \, 
|d(g)|^2 
\notag
\\
& \hspace{3ex}
\times 
\Theta\bigl(g\sinh\beta - |(\Omega/\M) + g f(g) \cosh\beta | \bigr) , 
\label{eq:Fcal-gen}
\end{align}
where we have introduced 
$K = |\k|$ and 
$g = K/\M$, 
and $\Theta$ is the Heaviside function. 
Note that $\M$ enters \eqref{eq:Fcal-gen} only as the overall 
factor and via the combination $\Omega/\M$. 

The crucial issue in \eqref{eq:Fcal-gen} is the behaviour of the argument of~$\Theta$: 
for what values of $\Omega$ is the argument of $\Theta$ positive for at least some interval of~$g$? 
If $f(x)\ge1$ for all~$x$, $\mathcal{F}(\Omega)$ clearly vanishes for all positive~$\Omega$: 
the detector does not become spontaneously excited. 
This is the case for the usual massless scalar field, 
for which 
$f(x)=1$, $d(x) = 2^{-1/2}$, and  
$\mathcal{F}(\Omega) = - \Omega \, \Theta(-\Omega)/(2\pi)$ \cite{birrell-davies}. 

We now specialize to $f$ and $d$ for which 
$f(x)\to1$ and $|d(x)|\to 2^{-1/2}$ as $x\to0_+$, 
so that the low energy dispersion relation 
and the low energy density of states reduce to those 
of the usual massless scalar field. 
Crucially, we assume that $f$ dips somewhere below unity. 
For concreteness, we further assume that 
$f_c := \inf f$ is positive, so that $0<f_c<1$. 
 Finally, we assume for simplicity that $d$ is everywhere nonvanishing. 
Under these quite broad assumptions, we now show that 
$\mathcal{F}(\Omega)$ has markedly different 
properties for rapidities below and above 
the critical value $\beta_c := \artanh(f_c)$.

Suppose first that $0 < \beta < \beta_c$. 
From the argument of $\Theta$ in \eqref{eq:Fcal-gen}
we see that $\mathcal{F}(\Omega)$ vanishes for $\Omega>0$ but not for $\Omega<0$: 
the detector does not become spontaneously excited, but it has a nonvanishing 
de-excitation rate. This is similar to the ordinary massless scalar field. 
The de-excitation rate is not Lorentz invariant, but at 
small negative values of $\Omega$ we find 
\begin{align}
\mathcal{F}(\Omega) 
& = - \frac{\Omega}{2\pi}
\biggl\{1 + \cosh\beta 
\Bigl[\bigl(1+\cosh(2\beta)\bigr) f'(0) 
\notag\\
& \hspace{12ex}
- 2 \Realpart\bigl(d'(0)/d(0)\bigr)\Bigr] h \, 
+ O(h^2)
\biggr\} , 
\label{eq:gen-smallbeta-smallgap}
\end{align}
where $h = \Omega/\M$, which shows that the Lorentz violation 
is suppressed at low energies by the factor~$\Omega/\M$. 
At $\M\to\infty$ with fixed~$\Omega$, 
\eqref{eq:gen-smallbeta-smallgap} reduces to the de-excitation rate of a detector 
coupled to the usual massless scalar field~\cite{birrell-davies}. 
This is all as one might have expected.

Suppose then that  $\beta > \beta_c$. Writing again $h = \Omega/\M$, 
the argument of $\Theta$ in \eqref{eq:Fcal-gen}
shows that $\mathcal{F}(\Omega)$ is now nonvanishing for $0 < h < 
\sup_{g\ge0} g[\sinh\beta - f(g) \cosh\beta]$: 
the detector gets spontaneously excited, at arbitrarily small positive $\Omega$! 
$\mathcal{F}(\Omega)$ has a nonvanishing limit as $\Omega\to0_+$, 
and the value of this limit is proportional to $\M$ 
by a function that depends only on~$\beta$. 

Similar observations for $\Omega<0$ 
show that $\mathcal{F}(\Omega)$ has a nonvanishing limit as $\Omega\to0_-$, 
and the value of this limit is again proportional to $\M$ 
by a function that depends only on~$\beta$. 

We summarize the general result: Suppose that the dispersion relation 
\eqref{eq:disprel-gen} is such that
the smooth positive-valued function $f$ on the positive real line satisfies 
$f(x)\to1$ as $x\to0_+$, and $f_c := \inf f$ satisfies $0<f_c<1$. 
Suppose further that the density of states 
\eqref{eq:density-gen} is such that the smooth complex-valued function 
$d$ on the positive real line satisfies $|d(x)|\to 2^{-1/2}$ as $x\to0_+$ and is nowhere vanishing. 
Then, an inertial UDW detector with rapidity $\beta >\beta_c=\artanh(f_c)$ in the preferred frame, experiences 
spontaneous excitations and de-excitations at arbitrarily low~$|\Omega|$,
and these transitions occur at a rate proportional to the Lorentz 
violation energy scale~$\M$.

\textit{\bfseries{Example: polymer quantum field theory.\thinspace---}}
LQG is one  of the approaches to quantum gravity \cite{Rovelli-book,Thiemann-book} currently being studied, with significant application to cosmology \cite{Bojowald:2001xe, Ashtekar:2006rx} (for reviews see \cite{Bojowald:2008zzb,Ashtekar:2015dja}). A key feature of the LQG formalism is an alternative quantization method called polymer quantization. This method has been applied, beyond its quantum gravity origins, to mechanical systems and to the scalar field \cite{Ashtekar:2002sn,Ashtekar:2002vh,Husain:2010gb,Hossain:2010eb,Hossain:2009ru,Seahra:2012un}. It is expected that for matter coupled to gravity in LQG, matter fields too are to be quantized using this prescription.  Hence it is natural to apply our general result on the excitation of inertial UDW  detectors to polymer quantization.  

We choose the specific implementation of a polymer quantized scalar field studied in~\cite{Hossain:2010eb}, 
referring therein for the details, and summarize here the features and results needed for our analysis. 
It has been previously observed \cite{Kajuri:2015oza} 
that an inertial detector coupled to a polymer quantized scalar field 
can become spontaneously excited: we shall 
establish both the critical rapidity for this to happen and the magnitude of the effect.

The Wightman function is given by 
\begin{align}
& G(t, \x; t', \x') 
= \int\frac{d^3\k}{(2\pi)^3} \, 
e^{i {\k}\cdot(\x-\x')}
\notag
\\
& \hspace{5ex}
\times \sum_{n=0}^\infty 
\bigl|c_{4n+3}(|\k|)\bigr|^2 e^{ -i \Delta E_{4n+3} (|\k|) 
(t-t'- i \epsilon)} , 
\label{eq:KGPropagator}
\end{align}
where 
$\Delta E_{4n+3}(|\k|) = |\k| f_{4n+3}(g)$, 
$c_{4n+3}(|\k|) = |\k|^{-1/2} d_{4n+3}(g)$,  
$g = |\k|/\M$, and the functions 
$f_{4n+3}$ 
and $d_{4n+3}$ may be expressed in terms of functions 
that appear in the theory of Mathieu's equation. 
The asymptotic expressions at small argument are 
\begin{subequations}
\label{eq:infrared-all}
\begin{align}
f_{4n+3}(g) &=  
\bigl[ (2n+1) - \tfrac12 (n+1)(2n+1) g 
\notag
\\
& \hspace{3ex}
- \tfrac{1}{16}(2n+1)(4n^2+7n+4) g^2 
+ O(g^3) \bigr] , 
\label{eq:e-infrared}
\\
d_3(g) &= 
\frac{i}{\sqrt{2}} \left[1 - \tfrac{3}{4} g - \tfrac{15}{32} g^2 
+ O(g^3) \right], 
\label{eq:d3-infrared}
\\
\frac{d_{4n+3}(g)}{d_{3}(g)} &= O(g^n) , 
\label{eq:cratio-infra}
\end{align}
\end{subequations}
while those at large argument are 
\begin{subequations}
\label{eq:ultraviolet-all}
\begin{align}
f_{4n+3} (g) &=  2(n+1)^2 \, g \left[ 1 + O(g^{-4}) \right],
\label{eq:e-ultraviolet}
\\
d_{3}(g) &= \frac{i}{4}\sqrt{\frac{g^3}{2}} 
\left[ 1 + O(g^{-4}) \right], 
\\
\frac{d_{4n+3}(g)}{d_{3}(g)} &= O(g^{-2n}) . 
\end{align}
\end{subequations}
If the polymer quantization is viewed as 
coming from an underlying quantum theory of spacetime, 
the polymer energy scale $\M$ may be identified as Planck energy. 

The detector's response $\mathcal{F}(\Omega)$ is obtained by applying formula 
\eqref{eq:Fcal-gen} to each $n$ in~\eqref{eq:KGPropagator},  
with $f(g) = f_{4n+3}(g)$ and $d(g) = d_{4n+3}(g)$, and summing over~$n$. 
From \eqref{eq:infrared-all} and \eqref{eq:ultraviolet-all} we see that 
$f_3(g)\to1$ as $g\to0_+$, 
$f_3(g)<1$ at small~$g$, $f_3(g)>1$ at large~$g$, and 
$|d_3(g)|\to 2^{-1/2}$ as $g\to0_+$; further, 
numerical experiments show that $f_3(g)$ dips below unity for 
$0 < g < g_m \approx 0.4334$, with a unique 
global minimum at $g = g_c \approx 0.2585$, 
such that $f_3(g_c) \approx 0.8781$. 
The $n=0$ contribution to $\mathcal{F}(\Omega)$ is 
hence precisely of the type analyzed above, with 
$\beta_c := \arctanh[f_3(g_c)] \approx 1.3675$, exhibiting a drastic low energy 
Lorentz violation for $\beta>\beta_c$. 
For $n>0$, 
numerical experiments indicate that $f_{4n+3}(g)>1$ for all~$g$, 
consistently with the asymptotic expressions in 
\eqref{eq:infrared-all} and \eqref{eq:ultraviolet-all}. 
This shows that the $n>0$ terms in 
\eqref{eq:KGPropagator} do not contribute to $\mathcal{F}(\Omega)$ for $\Omega>0$, 
and while these terms do contribute for $\Omega<0$, 
the expansions \eqref{eq:infrared-all} and \eqref{eq:ultraviolet-all} 
show that their effect is subdominant when $-\Omega/\M$ is small. 
At small and large negative values of $\Omega$ we obtain respectively 
\begin{align}
\mathcal{F}(\Omega) = - \frac{\Omega}{2\pi}
\bigl[1 - 2 (\cosh\beta \sinh^2\!\beta) \, h 
+ O(h^2)
\bigr]
\label{eq:smallbeta-smallgap}
\end{align}
and 
\begin{align}
\mathcal{F}(\Omega) = 
\frac{\M \sqrt{\cosh\beta}}{32\sqrt{2} \, \pi}
{(-h)}^{-3/2} 
\bigl[1 + O (h^{-2})
\bigr] , 
\label{eq:smallbeta-largegap}
\end{align}
where again $h = \Omega/\M$, and the only term in \eqref{eq:KGPropagator}   
that contributes to the order shown is the $n=0$ term.


\begin{figure}
 \begin{center}
  \includegraphics[height=6.5cm,width=\columnwidth]{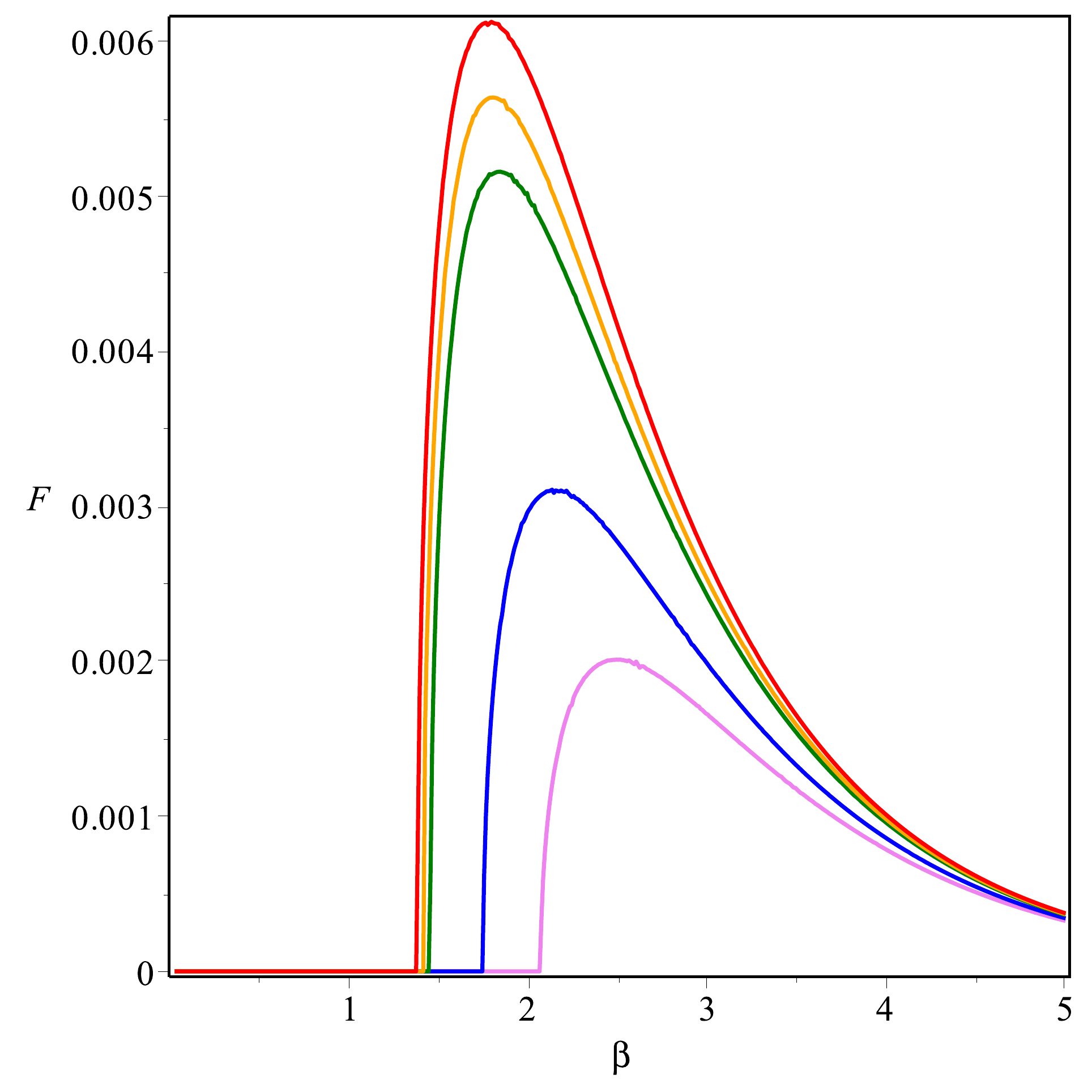}
 \end{center}
 \vspace{-5mm}
\caption{Transition rate for excitations ($h>0$) vs.\ 
$\beta$ for $h=0.1$ (lowest curve), $0.05$, $0.01$, $0.05$ and $0.01$ (highest curve).} 
\vspace{-3mm}
\label{fig:Fdot-beta-h+ve}
\end{figure}


\begin{figure}
 \begin{center}
  \includegraphics[height=6.5cm,width=\columnwidth]{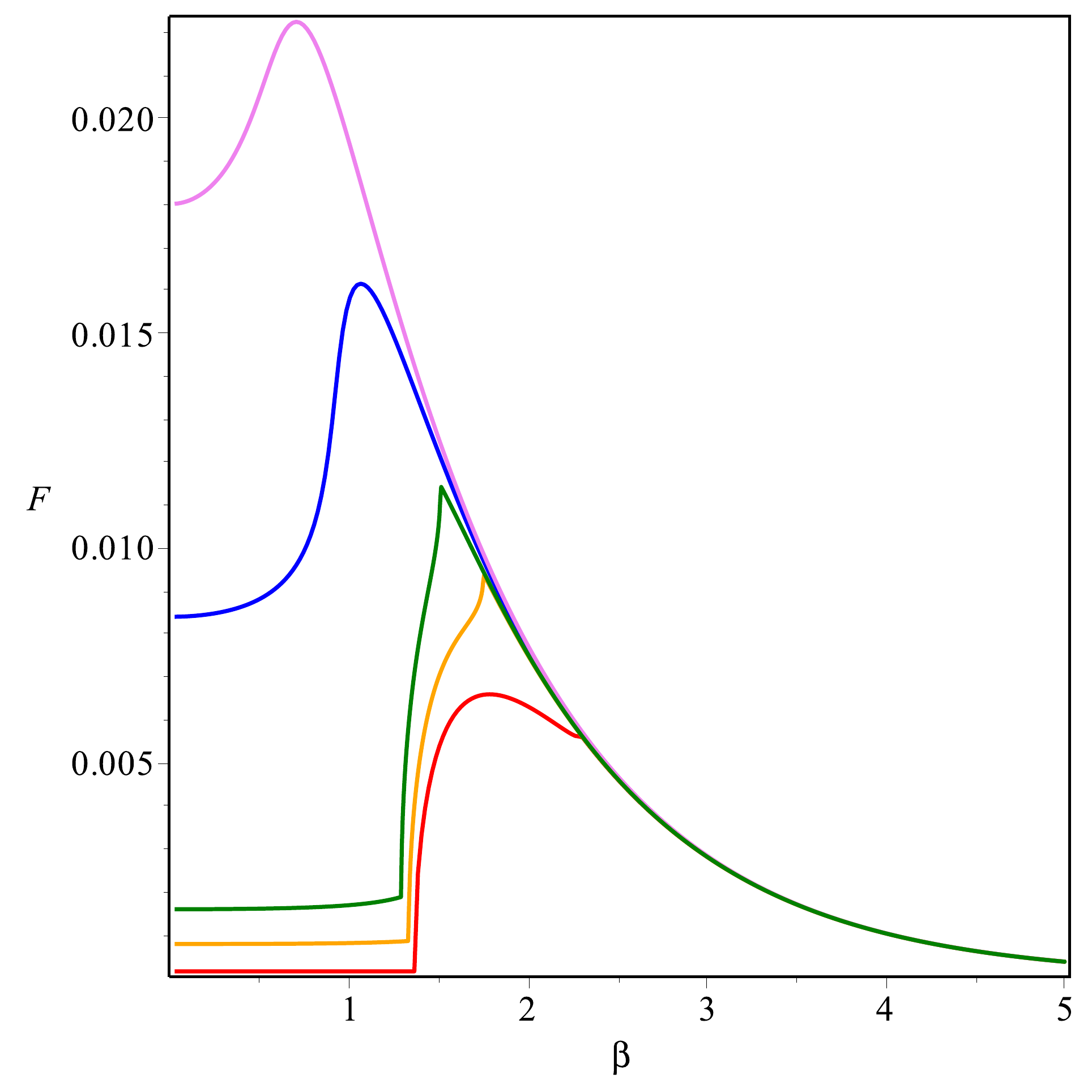}
 \end{center}
 \vspace{-5mm}
\caption{Transition rate for de-excitations ($h<0$) vs.\ 
$\beta$ for  $h=-0.1$ (highest curve), 
$-0.05$, $-0.01$,  $-0.005$  and $-0.001$ (lowest curve).}
 \vspace{-3mm}
\label{fig:Fdot-beta-h-ve}
\end{figure}


\begin{figure}
 \begin{center}
\includegraphics[height=6.5cm,width=\columnwidth]{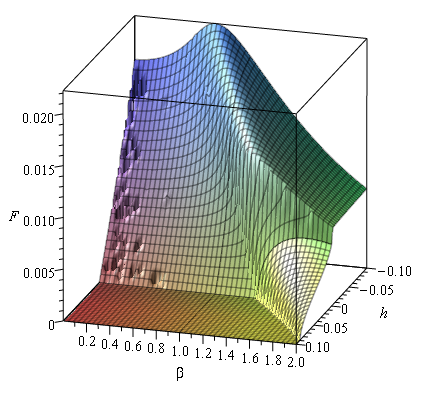}
 \end{center}
 \vspace{-5mm}
\caption{Transition rate
as a function of $\beta$ and~$h$. 
The irregularities at small $\beta$ are numerical noise.} 
\vspace{-3mm}
\label{fig:F-3d-1to2}
\end{figure}


Figures \ref{fig:Fdot-beta-h+ve}--\ref{fig:F-3d-1to2} 
show a numerical evaluation of 
$\mathcal{F}(\Omega)$, expressing $\Omega$ in terms of the dimensionless 
parameter~$h$. The scale of the $\mathcal{F}$-axis is in units of~$\M$. 
The plots use the $n=0$ term in in~\eqref{eq:KGPropagator}. 
For $h>0$ the $n=0$ term is the full contribution. 
For $h<0$ the $n=0$ term is a lower bound that becomes sharp as $h\to0_-$. 
The dramatic increase in the transition rate as $\beta$ increases past $\beta_c$ 
is apparent in all the plots. 

\textit{\bfseries{Discussion.\thinspace---}}
We have shown that an inertial UDW detector coupled to a scalar 
field with  a dispersion relation that violates Lorentz invariance at high energies, 
in a certain controlled way, exhibits drastic Lorentz violation at arbitrarily 
low energies. This violation occurs whenever the detector's velocity in the distinguished frame 
exceeds a certain critical rapidity~$\beta_c$, 
which is independent of the Lorentz violating energy scale~$\M$. 

The class of theories in which this phenomenon occurs includes 
a scalar field quantized in the polymer framework 
that imports techniques from loop quantum gravity. For the polymer 
quantized scalar field we  find $\beta_c \approx 1.3675$, 
independently of the polymer mass scale. 

The signature of the Lorentz violation is a sudden increase in the
rate of spontaneous de-excitations and a sudden emergence of
spontaneous excitations in the detector. Our linear perturbation theory analysis
predicts for these transition rates a magnitude comparable to the 
Lorentz-breaking energy scale~$\M$. 

As we have assumed the interaction to last much longer than~$1/\M$, 
and since transition probabilities
can by definition not be larger than unity, the predicted value for
the transition rate should not be taken literally: what the perturbative result
means is that the transition probabilities grow quickly to order
unity, after which a quantitative analysis would 
need to incorporate the back-reaction of 
the detector on the state of the
quantum field~\cite{hu-lin-louko,DeBievre:2006px}. 
Qualitatively, however, we view the linear perturbation theory result
as a reliable indicator of a violent burst of excitations and
de-excitations when the detector's rapidity exceeds~$\beta_c$.

Since the onset of excitations and the increase in the de-excitations 
occur when the detector's velocity exceeds the
group velocity of high frequency waves, 
the phenomenon can be compared to the the Cerenkov effect.
The surprise is that the effect shows up already at 
arbitrary small values of the detector's energy gap, much below the energies at 
which the field's group velocity is less than the speed of light.

While it may seem paradoxical that a symmetry violation
at high energies can induce large low energy effects, we emphasize that instances of this 
kind are known to occur in quantum field theory \cite{Polchinski:2011za,Collins:2004bp,bachmann-kempf}.

The close similarity between an UDW detector and the dipole moment interaction by which an
atom couples to the quantized electromagnetic field
\cite{PhysRevD.87.064038,PhysRevA.89.033835} suggests that our
prediction should apply to atoms or ions moving with a relativistic
velocity, including the ions accelerated to rapidity $\beta\approx3$
at the RHIC\null. The absence of observed Lorentz violations at the 
RHIC therefore provides 
a strong constraint on the possible families of modified dispersion relations.

Finally, although we have considered only an inertial detector, 
we anticipate that a uniformly linearly accelerated detector will exhibit a similar drastic 
Lorentz violation after operating so long that its rapidity in the preferred 
frame exceeds~$\beta_c$. 
It is therefore curious that the Hawking effect appears to be robust to high 
energy modifications of the dispersion relation~\cite{Unruh:1994je}, 
but the response of a uniformly accelerated UDW detector, 
which in some ways is its equivalence principle dual, is not so immune.

\smallskip
\noindent
\textit{\bfseries{Acknowledgments.\thinspace---}} 
We thank Achim Kempf for bringing Ref.\ \cite{bachmann-kempf} 
to our attention. 
V.H. was supported
by the Natural Science and Engineering Research Council of
Canada.  J.L. was supported in part by
STFC (Theory Consolidated Grant ST/J000388/1). 
J.L. thanks the organizers of the Atlantic GR 2015 meeting, 
supported by the Atlantic Association for Research in the Mathematical Sciences, 
for hospitality at the University of New Brunswick 
in the initial phase of this work.
 
\vfill

\bibliography{poly-detector}
\end{document}